\documentclass[11pt]{article}
\usepackage{graphicx}
\usepackage[margin=1.25in]{geometry}
\usepackage[usenames,dvipsnames]{color}
\usepackage{url}
\usepackage[colorlinks = true,
            linkcolor = blue,
            urlcolor  = blue,
            citecolor = blue,
            anchorcolor = blue]{hyperref}

\newcommand\pubnumber{Transcendental Preprint }
\newcommand\pubdate{\today}

\def\Title#1{\begin{center} {\LARGE #1 } \end{center}}
\def\Author#1{\begin{center}{ \sc #1} \end{center}}

\newcommand\pubblock{\rightline{\begin{tabular}{l} \pubnumber\\
         \pubdate \end{tabular}}}
\newenvironment{Abstract}{\begin{quotation} \begin{center}
                       ABSTRACT
     \end{center}\bigskip  }{\end{quotation}}

\def\beq{\begin{equation}}
\def\eeq#1{\label{#1}\end{equation}}
\def\eeqn{\end{equation}}

\newenvironment{Eqnarray}%
   {\arraycolsep 0.14em\begin{eqnarray}}{\end{eqnarray}}
\def\beqa{\begin{Eqnarray}}
\def\eeqa#1{\label{#1}\end{Eqnarray}}
\def\eeqan{\end{Eqnarray}}

\let\bar=\overbar

\def\lsim{\mathrel{\raise.3ex\hbox{$<$\kern-.75em\lower1ex\hbox{$\sim$}}}}
\def\gsim{\mathrel{\raise.3ex\hbox{$>$\kern-.75em\lower1ex\hbox{$\sim$}}}}

\def\del{\partial}
\def\Dslash{\not{\hbox{\kern-4pt $D$}}}
\def\dslash{\not{\hbox{\kern-2pt $\del$}}}
\def\pslash{\not{\hbox{\kern-2pt $p$}}}
\def\ETmiss{\not{\hbox{\kern-4pt $E$}}_T}

\def\Dlr{\mathrel{\raise1.5ex\hbox{$\leftrightarrow$\kern-1em\lower1.5ex\hbox{$D$}}}}

\def\MSB{{\bar{M \kern -2pt S}}}
\def\msb{{\bar{\scriptsize M \kern -1pt S}}}

\def\drb{{\bar{\scriptsize D \kern -1pt R}}}

\newcommand\snowmass{\begin{center}\rule[-0.2in]{\hsize}{0.01in}\\\rule{\hsize}{0.01in}\\
\vskip 0.1in Submitted to the  Proceedings of the US Community Study\\ 
on the Future of Particle Physics (Snowmass 2021)\\ 
\rule{\hsize}{0.01in}\\\rule[+0.2in]{\hsize}{0.01in} \end{center}}

\begin{document}

\pubblock

\Title{Particle Physics Outreach at Non-traditional Venues}

\bigskip 

\Author{Jim Cochran, John Huth, Roger Jones, Paul Laycock, Claire Lee, Lawrence Lee, Connie Potter, Gordon Watts}

\medskip


\medskip

 \begin{Abstract}
Since 2016 the group known as `The Big Bang Collective' has brought High Energy Physics outreach to music and culture festivals across Europe, successfully engaging with and inspiring audiences who may never have sought this out themselves through activities in their `Physics Pavilions'. The US has a very long, glorious tradition of music festivals, and an incredible community of sci-comm engaged physicists, from students to senior staff. With the experience already gained in Europe, The Big Bang Collective believes the time is now ripe to bring their Physics Pavilions to US music festivals.

\end{Abstract}

\snowmass

\def\TBBC{\emph{The Big Bang Collective}}

\def\thefootnote{\fnsymbol{footnote}}
\setcounter{footnote}{0}
\section{Introduction}
Particle physics outreach has seen a huge boost over the past decade, with large numbers of people engaging with scientists and institutions through social media, visits to facilities, dedicated outreach events, and more. We have now reached the era where these are ``traditional'' outreach methods, interacting with and educating people who are already interested in this type of science. This is why \TBBC{}'s music and culture festival program is such an innovation - they have put particle physics in front of tens of thousands of people who would \emph{not} normally seek to engage with science. Since 2016 Connie Potter (CERN), Roger Jones (Lancaster University), and Chris Thomas (Iowa State University) have organised and run large science installations in many of the best known music and culture festivals across Europe and the UK. Their efforts have introduced general science, particle physics, and the work and research done at CERN to a diverse group of people who would not normally have sought it out, with very exciting results. 
\\
\\
Participating in a festival programme is a unique way of reaching a public that generally has little direct contact with physics or science. Potter, Jones, and Thomas formed \TBBC{} as a way to take science to ‘where the people are’, in a fixed location (often a field!), and have worked to put scientific activities on the programme in a dedicated ``Physics Pavilion'' alongside musical concerts, art and literature events, comedy, children’s activities and other cultural experiences offered at the festival. As a result they have not only been able to engage with people in a meaningful way (a constant presence over a 3 or 4 day period, return visitors, and continuing conversations) but have also brought science into the mainstream consciousness, thereby breaking down the all-too-frequent misconception of physics and science as only ‘for the smart people’ or ‘for nerds’ or quite simply ‘not for me because I just don’t understand it’.\\
\\
The experience of \TBBC{}'s members showed that the key to success at these festivals is tailoring the programme to the demographic of the audience and the ‘style’ of the festival itself, together with having a good team of volunteer physicists of all ages on hand. People attending these events are often in awe of the fact that they have actual scientists in front of them and are able to ask them any questions they like. They are relaxed and happy in the festival environment, which is the ideal situation for learning to take place. Even those festival-goers who really did not like science at school, gave up on it and ever since have steered away from all things ‘science’, have had their curiosity piqued and begun engaging with the subject again. Many have become repeat visitors, returning each year to see what the Physics Pavilion has in store this time.

\begin{figure}[tb]
    \centering
    \frame{\includegraphics[width=0.65\textwidth]{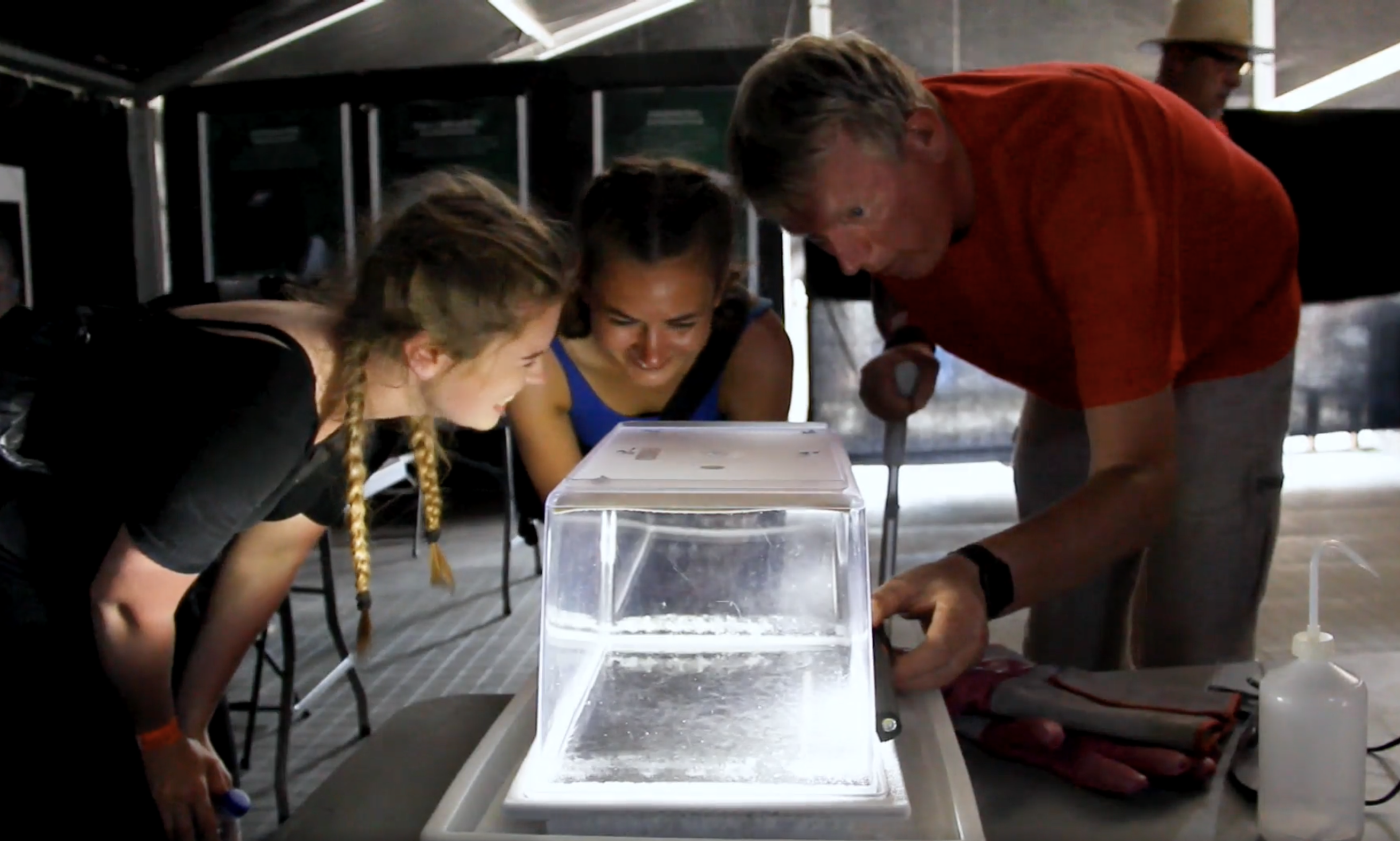}}
    \caption{The ``Build Your Own Cloud Chamber'' workshop at Roskilde Festival 2019.~\cite{Video_Roskilde}
    }
    \label{fig:cloudChamber}
\end{figure}

\section{Format and Event History}

Each Physics Pavilion is a collaboration between CERN, the festival itself and the physics departments of institutes or universities in the local area. Responsibility for resources and finance is shared between the three partners and managed by the core team.\\

\begin{figure}[tb]
    \centering
    \frame{\includegraphics[width=0.65\textwidth]{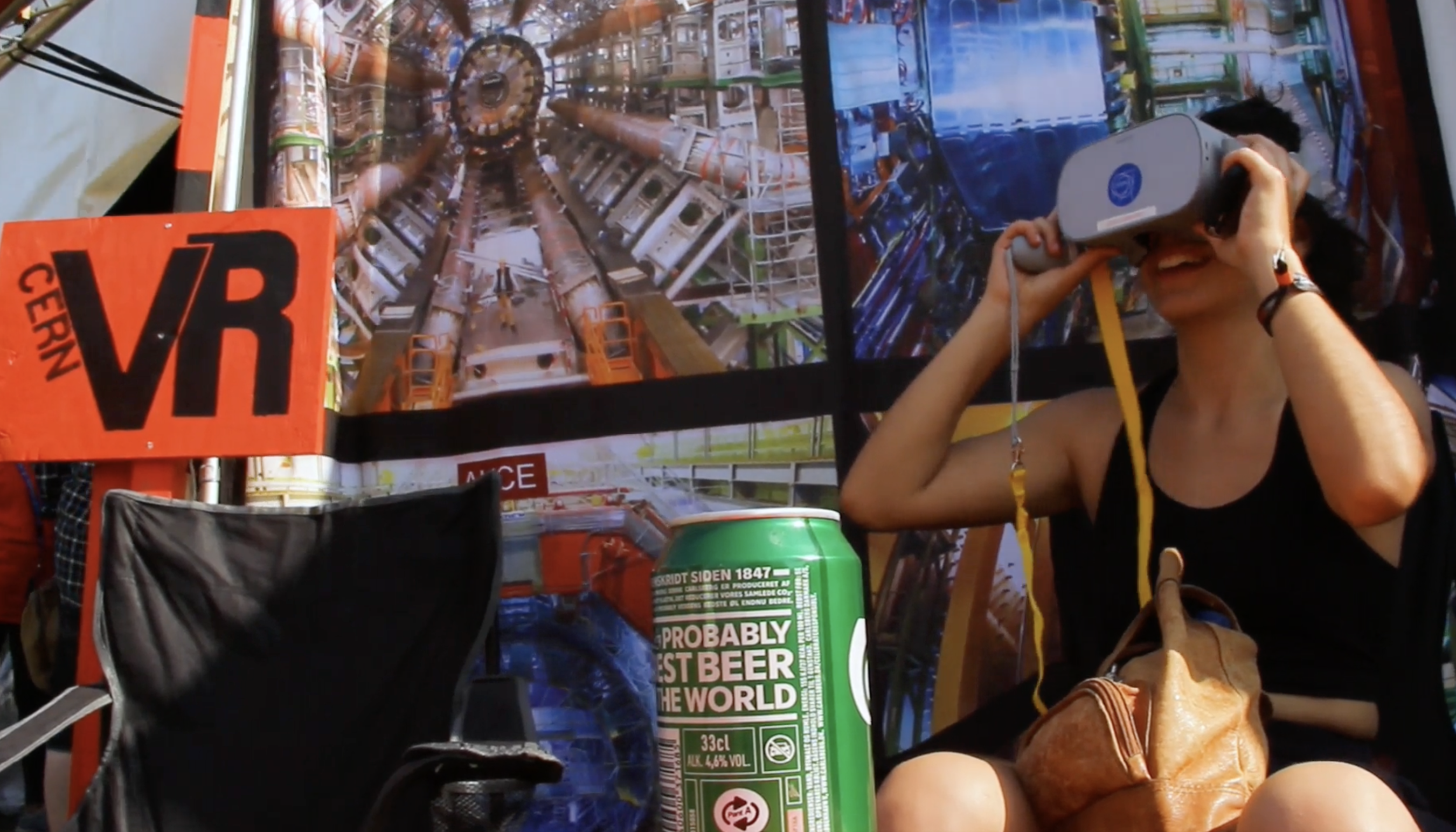}}
    \caption{The CERN Virtual Reality station at Roskilde Festival 2019~\cite{Video_Roskilde}.
    }
    \label{fig:vr}
\end{figure}

The general Physics Pavilion format stretching over the entire festival of 3 or more days includes a curated programme of talks, dedicated workshops such as “Build Your Own Cloud Chamber” (shown in Figure~\ref{fig:cloudChamber}), a live virtual link-up (sometimes to CERN, sometimes NASA), and additional outdoor hands-on activities available continuously throughout the day, such as a Virtual Reality tour of LHC experiments (Figure~\ref{fig:vr}). The speakers and topics provide a variety of content along three broad general themes: CERN speakers on CERN-related research, national speakers on current hot topics in HEP, and well-known science personalities on more general science topics.\\
\\
The first Physics Pavilion was hosted in 2016 at the international WOMAD~\cite{WOMAD} festival in the UK, and has been an integral part of the festival ever since, growing in size with each consecutive year. In 2019, \TBBC{} expanded to bring their Physics Pavilions to a total of four music festivals across Europe: WOMAD (UK)~\cite{WOMAD2019}, Colours of Ostrava (Czech Republic)~\cite{Ostrava2019}, Pohoda (Slovakia)~\cite{Pohoda2019} and Roskilde (Denmark)~\cite{Roskilde2019}.\\
\\
All festivals take place in what is known as ‘festival season’ (European summer). In 2020 all four Pavilions were on schedule until the arrival of Covid-19, however, the WOMAD Physics Pavilion ran virtually with a series of talks still available on YouTube~\cite{Video_WOMAD2020}. \TBBC{} is looking forward to resuming the live events once the pandemic is over and events can safely take place once again.

\section{Impact}

The Physics Pavilions have had a tremendous impact on festival-goers of all ages. Younger audience members have been enthralled by the hands-on activities, which have often awakened an interest in science and technology. Parents have discovered that their child has a talent for science or computing that they may not have had a chance to nurture before. Volunteers have been asked questions from parents asking what topics their child should study at school as they realise how much they enjoy science. Feedback from adults has been hugely positive as well, with many people reiterating what an excellent idea it is to have science at a music festival. And of course, the ultimate proof of concept is the number of visitors who return to the Physics Pavilion year after year.\\
\\
To get a good feel for the excitement, energy and inspiration the Physics Pavilions provide, highlight videos from WOMAD 2018~\cite{Video_WOMAD2018}, 2017~\cite{Video_WOMAD2017}, 2016~\cite{Video_WOMAD2016}, and Roskilde 2019~\cite{Video_Roskilde} are available online.

\section{Extension to the United States}

The United States has a long history of music festivals. From Newport, Monterey, and Woodstock to the more contemporary Lollapalooza and SXSW festivals, statistics from Billboard~\cite{Billboard} indicate that in 2014, 32 million people attended at least one U.S. music festival. Nielsen Music reports that the average attendee travels approximately 903 miles to attend a festival. In an article in TIME magazine on the rise in popularity of music festivals, it was noted that ‘people are more likely now to spend money on experiences over material goods’~\cite{Time}. That ‘experiential economy’ works in favour of including other types of ‘experiences’ on a festival’s programming. Introducing particle physics events, such as a ‘build your own cloud chamber’ workshop, can uniquely expand the breadth of a festival's offerings.

Two new LHC faculty at the University of Tennessee, Knoxville, Profs. Tova Holmes and Lawrence Lee, aspire to connect with the Bonnaroo Music and Arts Festival held each year on a 700-acre farm in Manchester, Tennessee. In ordinary years, Bonnaroo welcomes more than 80,000 visitors and is one of the largest music festivals in the US. Bringing \TBBC{}'s existing large-festival programming to Bonnaroo would be a natural fit. Lee's \textsc{ColliderScope} project presents a live musical show where wave forms draw images from the world of particle physics. The project has served as a musical element of the Physical Pavilions starting in 2019. Live shows at the festival stages at the Pohoda Festival (2019) and Roskilde (2022) marry the musical program of the festivals with our scientific program. Now local to Tennessee, \textsc{ColliderScope} can be easily incorporated into a potential Bonnaroo programme.

Prof. Gordon Watts and the group at the University of Washington, Seattle would like to incorporate the \TBBC{}'s program into festivals in Seattle such as the Northwest Folklife Festival and Bumbershoot, both festivals that welcome well over 100,000 visitors each year.

The breadth of these American music festivals is such that each comes with a unique community that can be reached with our programming to get them excited about physics. A folk festival and an EDM festival have different slices of the public to reach, and in the spirit of reaching new audiences, we advocate for partnering with a wide range of festivals.

\section{Conclusion}
The Physics Pavilion initiative by \TBBC{} has shown to be an extremely successful, innovative, and inspiring effort in bringing science education and outreach to music and culture festivals over the past 6 years. The success of this programme can and should be extended to music festivals in the United States in the coming years.

\bibliographystyle{unsrt}
\bibliography{references}

\begin{thebibliography}{10}

\bibitem{Video_Roskilde}
{Roskilde 2019 Physics Pavilion Highlights}.
\newblock \url{https://youtu.be/L7U0K4nww6E}.

\bibitem{WOMAD}
{WOMAD World of Music, Arts and Dance}.
\newblock \url{https://womad.co.uk/}.

\bibitem{WOMAD2019}
{WOMAD World of Physics 2019}.
\newblock \url{https://womad.co.uk/the-world-of-physics-2019-line-up/}.

\bibitem{Ostrava2019}
{Colours of Ostrava Festival Big Bang Stage 2019}.
\newblock \url{http://ipnp.cz/thebigbangstage/}.

\bibitem{Pohoda2019}
{Pohoda Festival Science Stage 2019}.
\newblock
  \url{https://www.pohodafestival.sk/en/news/cern-comenius-university-and-the-big-bang-collective-present-magical-science-at-pohoda-2019}.

\bibitem{Roskilde2019}
{Roskilde Festival Physics Pavilion 2019}.
\newblock
  \url{https://www.roskilde-festival.dk/en/years/2019/acts/cern-and-the-niels-bohr-institute/}.

\bibitem{Video_WOMAD2020}
{Virtual WOMAD 2020}.
\newblock
  \url{https://youtube.com/playlist?list=PLAV2PYKNpJERPTSorpuxDj320Bo1tEXoi}.

\bibitem{Video_WOMAD2018}
{WOMAD 2018 Physics Pavilion Highlights}.
\newblock \url{https://vimeo.com/showcase/5373156/video/311515696}.

\bibitem{Video_WOMAD2017}
{WOMAD 2017 Physics Pavilion Highlights}.
\newblock \url{https://vimeo.com/286656947}.

\bibitem{Video_WOMAD2016}
{WOMAD 2016 Physics Pavilion Highlights}.
\newblock \url{https://vimeo.com/311512363}.

\bibitem{Billboard}
Billboard: Check out these surprising stats about u.s. music festivals.
\newblock
  \url{https://www.billboard.com/culture/events/music-festival-statistics-graphic-6539009/}.

\bibitem{Time}
Time: How music festivals became a massive business in the 50 years since
  woodstock.
\newblock \url{https://time.com/5651255/business-of-music-festivals/}.

\end{thebibliography}
\end{document}